\documentclass[10pt,journal,final]{IEEEtran}
\IEEEoverridecommandlockouts

\usepackage{amsfonts,amssymb}
\usepackage{ifpdf}
\usepackage{cite}
\usepackage{stfloats}
\usepackage{bm}
\usepackage{color,xcolor}
\usepackage{subfigure}
\ifCLASSINFOpdf
   \usepackage[pdftex]{graphicx}
   \graphicspath{{../pdf/}{../jpeg/}}
   \DeclareGraphicsExtensions{.pdf,.jpeg,.png}
\else
   \usepackage[dvips]{graphicx}

   \graphicspath{{../eps/}}

   \DeclareGraphicsExtensions{.eps}
\fi

\usepackage[cmex10]{amsmath}

\usepackage{algorithm}
\usepackage{algorithmic}
\ifCLASSOPTIONcompsoc
\else
  \usepackage[caption=false,font=footnotesize]{subfig}
\fi

\usepackage{makecell}
\begin{document}
\title{ Throughput   Maximization for   Full-Duplex UAV Aided Small Cell  Wireless Systems}
\author{Meng~Hua,~\IEEEmembership{Student Member,~IEEE,}
Luxi~Yang,~\IEEEmembership{Senior Member,~IEEE,}
Cunhua Pan,\\
and~Arumugam Nallanathan,~\IEEEmembership{Fellow,~IEEE}
\thanks{Manuscript received August    20, 2019; revised October     19, and accepted December   9, 2019. This work was supported by National Natural Science Foundation of China under Grant  61971128,  Grant 61372101, and  Scientific Research Foundation of Graduate School of Southeast University  under Grand  YBPY1859 and China Scholarship Council (CSC) Scholarship, National High Technology Project of China  under 2015AA01A703.   The associate editor coordinating the review of this paper and approving it for publication was Ayca Ozcelikkale. (\emph{Corresponding author: Luxi Yang}.)}
\thanks{M. Hua,   and L. Yang are with the School of Information Science and Engineering, Southeast University, 210096, China. (e-mail: \{mhua, lxyang\}@seu.edu.cn).
}
\thanks{C. Pan,  and A. Nallanathan are with the School of Electronic Engineering and Computer Science, Queen Mary University of London, London E1 4NS, U.K. (e-mail: \{c.pan, a.nallanathan\}@qmul.ac.uk)}

}
\maketitle
\begin{abstract}
This paper investigates full-duplex unmanned aerial vehicle (UAV) aided small cell wireless systems, where  the UAV  serving as the base station (BS)  is designed  to transmit data to the downlink users and receive data from the uplink users simultaneously. To maximize the total system capacity, including  uplink and downlink capacity,  the  UAV trajectory, downlink/uplink user scheduling, and uplink user transmit power are alternately  optimized. The resulting optimization problem is  mixed-integer and non-convex, which is challenging to solve. To address it, the block coordinate descent method  and successive convex approximation techniques are leveraged. Simulation  results demonstrate the significant capacity gain can be achieved by our design compared with the other designs.
\end{abstract}

\section{Introduction}
Recently,  unmanned aerial vehicles (UAVs) have received significant research interests both from academia and industry as a promising technique for various  applications  such as data collection, wireless power transfer,  hot-spot offloading, data transmission, etc, \cite{zhan2018energy,xu2018uav,lyu2018uav,wu2018Joint}.
A typical functionality of UAV is acted as a mobile base station (BS).  The authors in \cite{zhan2018energy} studied UAV-aided data collection problem with the objective of minimizing the maximum energy consumption of all sensors by jointly optimizing the communication access strategy and the UAV trajectory. The authors in \cite{xu2018uav} studied the single UAV-enabled multiuser wireless power transfer system that targets at maximizing the amount of energy transferred to the total users by optimizing the UAV trajectory. The hot-spot problem was addressed by \cite{lyu2018uav}, where the authors used the UAV  to cover cell-edge users and offload the data traffic  from the overloaded BS.  A multi-UAV enabled system for serving multiple users was presented  in \cite{wu2018Joint} to improve throughput by carefully designing the UAV trajectories and their transmit power. A sustainable UAV communication was  investigated in \cite{sun2019optimal}, where the authors proposed a solar-powered UAV to serve users  with energy harvested  from sun by adjusting its altitude and  horizontal trajectory. In addition, the UAV can also act as a relay. For example, work \cite{ZhangJoint2018} studied  the UAV-aided relay system, where the user communicated with BS with the help of  UAV   to minimize the system outage by optimizing the UAV trajectory and transmit power.

The full-duplex technique allows  the downlink and uplink transmission operating at the same time and frequency, and thus can  double the system capacity  compared with the  half-duplex technique \cite{choi2010achieving}. At present, there have been some work on the research of full-duplex UAV \cite{hua2018outage},\cite{spectrum2018wang}.  In  \cite{hua2018outage},  the authors considered the time-sensitive scenario, where the full-duplex UAV acts  as a relay to  minimize the relaying system outage probability. The authors in  \cite{spectrum2018wang} further considered a more complicated scenario, where  the full-duplex UAV serving in a   device-to-device underlaying celluar system  was studied. However, both of them  focus on studying the UAV relaying system, the  full-duplex UAV acting  as  mobile BS is still not investigated.

In this paper, we  deploy  a  full-duplex UAV-BS to serve the targeted small cell users, including uplink and downlink users. In  the uplink transmission phase, multiple uplink users transmit their data to the UAV with TDMA manner. Meanwhile,  the full-duplex UAV-BS transmits the data to  multiple downlink  users in the downlink transmission phase still with TDMA manner. However, the downlink users will receive strong interference from the uplink users.  Therefore, a fundamental question  for the proposed full-duplex UAV-BS enabled  systems is  how to jointly optimize the UAV trajectory, uplink user transmit power, downlink and uplink user scheduling   so as to maximize the system uplink and downlink capacity. To tackle this challenge,  we divide the resulting problem into four sub-problems and optimize one subset of variables while keeping other variables fixed, and then alternately optimize the four sub-problems in an iterative way  by using the block coordinate descent method and successive convex optimization techniques.  The numerical results demonstrate that the proposed design  significantly outperforms the benchmarks.

\begin{figure}[!t]
\centerline{\includegraphics[width=2.5in]{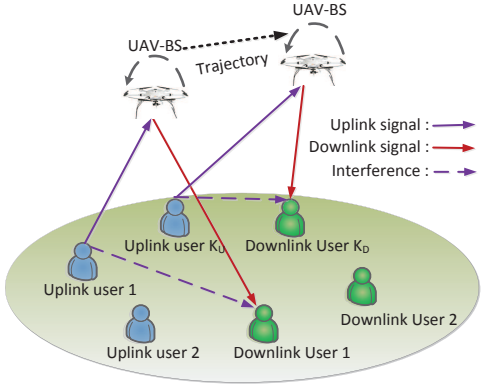}}
\caption{Full-duplex  UAV aided small cell wireless systems} \label{fig1}
\end{figure}
\section{System Model And Problem Formulation}
We consider a UAV-enabled communication system where the UAV serves as a full-duplex BS  that can communicate with $K_D$ single-antenna downlink users  and $K_U$ single-antenna uplink users  using the same and frequency resource as shown in Fig.~\ref{fig1}.  The UAV is equipped with two antennas, in which one is used for data transmission in the downlink  and the other is used for data reception in the uplink.  Define  ${{\cal K}_D}{\rm{ = }}\left\{ {1,2, \ldots ,{K_D}} \right\}$ and ${{\cal K}_U}{\rm{ = }}\left\{ {1,2, \ldots ,{K_U}} \right\}$ as the sets of downlink and uplink users, respectively. We denote the horizontal coordinate of user $k$ as ${\bf w}_k$, $k \in {{\cal K}_D} \cup {{\cal K}_U}$. The UAV altitude is fixed at $H$. The given  time period $T$ is equally  divided into $N$ time slots with duration $\delta=T/N$. Then, the horizontal location of UAV at any slot $n$ is  denoted as ${\bf q}[n]$.

As pointed out  by the 3GPP,  the UAV-ground channel model depends on the environmental scenarios, such as the suburban with less scattering and the macro urban with rich scattering. Especially when the UAV flies above $40 \rm m$ in the rural area,  the UAV offers a nearly 100\% LoS probability for  UAV-ground channel as shown  in the 3GPP specification \cite{3GPP}. As a consequence, the downlink channel  gain of  UAV to the $j\text{-}\rm{th}$ downlink user at time slot $n$ is given by \cite{power2018meng,zeng2017energy,spectrum2018wang,duo2019energy}

\begin{align}
h_{b,j}[n] = {\beta _0}{\left( {{{\left\| {{\bf{q}}\left[ n \right] - {{\bf{w}}_j}} \right\|}^2} + {H^2}} \right)^{ - 1}},j \in {{\cal K}_D},
\end{align}
where  $\beta_0$ represents the reference channel gain at $d =1{\rm m}$. Similarly,  the uplink channel gain from the $i\text{-}\rm{th}$ uplink user to the UAV at time slot $n$ is denoted as $h_{i,b}[n]$, $i \in {{\cal K}_U}$.

The channel model from the $i\text{-}\rm{th}$ uplink user to the  $j\text{-}\rm{th}$ downlink user follows Rayleigh fading with  channel power gain denoted by ${g_{i,j}[n]} = {\beta _0}d_{i,j}^{ - \alpha}\xi$, where $d_{i,j}$ is the distance between  the $i\text{-}\rm{th}$ uplink user and the  $j\text{-}\rm{th}$ downlink user, $\alpha$ denotes the path loss exponent, and $\xi$ is a random variable following the exponential distribution with unit mean.

We adopt a TDMA manner for both downlink  and uplink users, and  assume that  the UAV can only  communicate with at most one user at one time slot in the uplink/downlink \cite{wu2018Joint},\cite{power2018meng},  which yields the following user  scheduling constraints
\begin{small}
\begin{align}
&\sum\limits_{j = 1}^{{K_D}} {x_j^d\left[ n \right]}  \le 1,\forall n, \label{const1}\\
&x_j^d\left[ n \right] \in \left\{ {0,1} \right\},\forall j,n,\label{const2}\\
&\sum\limits_{i = 1}^{{K_U}} {x_i^u\left[ n \right]}  \le 1,\forall n,\label{const3}\\
&x_i^u\left[ n \right] \in \left\{ {0,1} \right\},\forall i,n,\label{const4}
\end{align}
\end{small}
where \eqref{const1}\text{-}\eqref{const2} denote the downlink scheduling constraints, and \eqref{const3}\text{-}\eqref{const4} denote the uplink scheduling constraints.

The lower bound for the downlink ergodic capacity  from the UAV to the $j\text{-}\rm{th}$ downlink user at time slot $n$ is given by
\begin{small}
\begin{align}
R_j^d\left[ n \right] = {\log _2}\left( {1 + \frac{{{p_b}{h_{b,j}}\left[ n \right]}}{{\sum\limits_{i = 1}^{{K_U}} {x_i^u\left[ n \right]{\beta _0}d_{i,j}^{ - \alpha }{p_i}\left[ n \right] + {\sigma ^2}} }}} \right), \label{section2downlink}
\end{align}
\end{small}
where $\sigma ^2$ denotes the noise power, and $p_b$ represents the UAV transmit power. Similarly, the uplink capacity from the  $i\text{-}\rm{th}$ uplink user to the UAV  is given by
\begin{align}
R_i^u\left[ n \right] = {\log _2}\left( {1 + \frac{{{p_i}\left[ n \right]{h_{i,b}}\left[ n \right]}}{{{f_b}\left[ n \right] + {\sigma ^2}}}} \right), \label{section2uplink}
\end{align}
where $f_b[n]$ denotes the self-interference at time slot $n$ from the transmit antenna to the receive antenna at the full-duplex UAV (Here, we assume that $f_b[n]$ is a constant to represent the maximal self-interference on the UAV \cite{spectrum2018wang}), and $p_i[n]$ denotes the $i\text{-}\rm{th}$ uplink user transmit power at time slot $n$.

Let ${{\bf{A}}_{\rm D}} = \left\{ {x_j^d\left[ n \right],\forall j,n} \right\},{{\bf{A}}_{\rm U}} = \left\{ {x_i^u\left[ n \right],\forall i,n} \right\},{\bf{P}} = \left\{ {{p_i}\left[ n \right],\forall i,n} \right\},{\bf{Q}} = \left\{ {{\bf{q}}\left[ n \right],\forall n} \right\}$. We aim at   maximizing  the total system capacity, including  uplink and downlink links capacity, which is formulated as follows
\begin{align}
\left( {\rm{P}} \right)&\mathop {\max }\limits_{{\bf{P}},{\bf{Q}},{{\bf{A}}_{\rm {D}}},{{\bf{A}}_{\rm {U}}}} \sum\limits_{n = 1}^N {\sum\limits_{j = 1}^{{K_D}} {x_j^d\left[ n \right]} } R_j^{d}\left[ n \right] + \sum\limits_{n = 1}^N {\sum\limits_{i = 1}^{{K_U}} {x_i^u\left[ n \right]} } R_i^u\left[ n \right]\notag\\
&{\rm s.t.}~~0 \le {p_i}\left[ n \right] \le {P_{\max }},\forall i,n,\label{const5}\\
&\quad \left\| {{\bf{q}}\left[ n \right] - {\bf{q}}\left[ {n - 1} \right]} \right\| \le {V_{\max }}\delta ,\forall n,\label{const6}\\
&\quad {{\bf{q}}_I} = {\bf{q}}\left[ 0 \right],{{\bf{q}}_F} = {\bf{q}}\left[ N \right],\label{const7}\\
&\quad \eqref{const1},\eqref{const2},\eqref{const3},\eqref{const4},\notag
\end{align}
where $P_{\rm max}$ and $V_{\rm max}$ respectively denote the maximum  uplink user transmit power and  UAV speed, ${\bf{q}}_I$ and ${\bf{q}}_F$ represent UAV's initial and final location, respectively.
\section{Proposed Algorithm}
The objective function is a function of $\bf Q$, $\bf P$, ${\bf A}_{\rm D}$, and ${\bf A}_{\rm U}$, which is not jointly  convex with these variables. In addition, the binary  constraints of \eqref{const2} and \eqref{const4} make the optimization more difficult to solve. The optimal solution is hard to obtain even using exhaustive search. First, the search space is  ${\cal O}\left( {{{\left( {{K_D}{K_U}} \right)}^N}} \right)$ for solving downlink/uplink user scheduling, which means the  complexity is exponentially increasing with the number of time slots $N$. Second, even with the  fixed downlink and uplink user scheduling, the sub-problem is still non-convex with respective to uplink user transmit power and UAV trajectory, which indicates that the optimal solution still can not be obtained.

To deal with these issues, we first relax the binary variables ${\bf A}_{\rm D}$ and ${\bf A}_{\rm U}$ into continuous ones, and transform the  binary constraints \eqref{const2} and \eqref{const4} into the linear constraints, which are respectively given by
\begin{align}
0 \le x_j^d\left[ n \right] \le 1,\forall j,n,\label{const2NEW}\\
0 \le x_i^u\left[ n \right] \le 1,\forall i,n.\label{const4NEW}
\end{align}
Then, we propose a four-stage iterative optimization algorithm for solving problem $(\rm P)$, and a local solution is obtained.
\subsection{Stage 1: Downlink user scheduling  design}
First, we consider the  downlink user scheduling problem with the given   uplink user scheduling ${\bf A}_{\rm U}$, transmit power $\bf P$, and UAV trajectory $\bf Q$. Then, Problem $\rm P$ becomes the following optimization problem
\begin{small}
\begin{align}
\left( {{\rm{P1}}} \right)&\mathop {\max }\limits_{{{\bf{A}}_{\rm{D}}}} \sum\limits_{n = 1}^N {\sum\limits_{j = 1}^{{K_D}} {x_j^d\left[ n \right]} } R_j^{d}\left[ n \right]\notag\\
&{\rm s.t.}~~\eqref{const2NEW}.\notag
\end{align}
\end{small}
Obviously,  problem $(\rm P1)$ is a standard linear programming problem, and can be efficiently solved by using standard optimization packages such as CVX.
\subsection{Stage 2: Uplink user scheduling  design}
Second, we study the uplink user scheduling problem with  the given transmit power $\bf P$,  downlink user scheduling ${\bf A}_{\rm D}$, and  UAV trajectory $\bf Q$, which can be formulated as
\begin{small}
\begin{align}
\left( {{\rm{P2}}} \right)&\mathop {\max }\limits_{{{\bf{A}}_{\rm{U}}}} \sum\limits_{n = 1}^N {\sum\limits_{j = 1}^{{K_D}} {x_j^d\left[ n \right]} } R_j^d\left[ n \right] + \sum\limits_{n = 1}^N {\sum\limits_{i = 1}^{{K_U}} {x_i^u\left[ n \right]} } R_i^u\left[ n \right]\notag\\
&{\rm s.t.}~~\eqref{const4NEW}.\notag
\end{align}
\end{small}
The term $R_j^d\left[ n \right]$ in objective function of problem  $(\rm P2)$ is a strictly convex with respect to (w.r.t.) $x_i^u[n]$ and hence   not concave. Hence, problem $(\rm P2)$ is a non-convex optimization problem. To deal with this issue, we approximate the convex function as its lower bound, which is linear function of the optimization variables that is much easier to solve.  By  taking the first-order Taylor expansion at any feasible point $\{x_i^{u,r}\}$, we have
\begin{small}
\begin{align}
&R_j^d\left[ n \right] \ge R_j^{d,lb}\left[ n \right] \overset{\triangle} {=} {\log _2}\left( {1 + \frac{{{p_b}{h_{b,j}}\left[ n \right]}}{{I_j^r\left[ n \right]}}} \right) -  \notag\\
&\quad\sum\limits_{i = 1}^{{K_U}} {\frac{{{\beta _0}d_{i,j}^{ - \alpha }{p_i}\left[ n \right]{p_b}{h_{b,j}}\left[ n \right]\log _2^e}}{{I_j^r\left[ n \right]\left( {I_j^r\left[ n \right] + {p_b}{h_{b,j}}\left[ n \right]} \right)}}\left( {x_i^u\left[ n \right] - x_i^{u,r}\left[ n \right]} \right),}
\end{align}
\end{small}
where $I_j^r\left[ n \right] = \sum\limits_{i = 1}^{{K_U}} {x_i^{u,r}\left[ n \right]{\beta _0}d_{i,j}^{ - \alpha }{p_i}\left[ n \right] + {\sigma ^2}}$.
As a result, for  any given local point $\{{x_i^{u,r}}\}$,  define the following problem
\begin{align}
\left( {{\rm{{\bar P}2}}} \right)&\mathop {\max }\limits_{{{\bf{A}}_{\rm{U}}}} \sum\limits_{n = 1}^N {\sum\limits_{j = 1}^{{K_D}} {x_j^d\left[ n \right]} } R_j^{d,lb}\left[ n \right] + \sum\limits_{n = 1}^N {\sum\limits_{i = 1}^{{K_U}} {x_i^u\left[ n \right]} } R_i^u\left[ n \right]\notag\\
&{\rm s.t.}~~\eqref{const4NEW}.\notag
\end{align}
Now, problem $(\rm {\bar P}2)$ can be readily shown to be a convex optimization problem that can be efficiently solved. Then, $(\rm { P}2)$  can be approximately
solved by successively updating the uplink user scheduling ${{{\bf{A}}_{\rm{U}}}}$  obtained from $(\rm {\bar P}2)$.

\subsection{Stage 3: UAV trajectory design}
Third, we study the UAV trajectory optimization problem with  the given transmit power $\bf P$, downlink  user scheduling ${\bf A}_{\rm D}$, and uplink user scheduling ${\bf A}_{\rm U}$, which is given by
\begin{align}
\left( {{\rm{P3}}} \right)&\mathop {\max }\limits_{\bf{Q}} \sum\limits_{n = 1}^N {\sum\limits_{j = 1}^{{K_D}} {x_j^d\left[ n \right]} } R_j^d\left[ n \right] + \sum\limits_{n = 1}^N {\sum\limits_{i = 1}^{{K_U}} {x_i^u\left[ n \right]} } R_i^u\left[ n \right]\notag\\
&{\rm s.t.}~~\eqref{const6},\eqref{const7}.\notag
\end{align}
Problem $(\rm { P}3)$ is a non-convex optimization problem since the objective function is non-convex. To handle the non-convex objective function, the successive convex approximation technique  is  applied. It can be observed that   $R_j^d\left[ n \right]$  is convex w.r.t. ${\left\| {{\bf{q}}\left[ n \right] - {{\bf{w}}_j}} \right\|^2}$, but it is not convex w.r.t. ${\bf q}[n]$. By taking the first-order Taylor expansion at any given local point ${\left\| {{{\bf{q}}^r}\left[ n \right] - {{\bf{w}}_i}} \right\|^2}$, we can obtain its convex lower bound as follows
\begin{small}
\begin{align}
&R_j^d\left[ n \right] \ge {\log _2}\left( {1 + \frac{{{C_j}\left[ n \right]}}{{{{\left\| {{{\bf{q}}^r}\left[ n \right] - {{\bf{w}}_j}} \right\|}^2} + {H^2}}}} \right)- {\Sigma _j}\left[ n \right]\times\notag\\
&\left( {{{\left\| {{\bf{q}}\left[ n \right] - {{\bf{w}}_j}} \right\|}^2} - {{\left\| {{{\bf{q}}^r}\left[ n \right] - {{\bf{w}}_j}} \right\|}^2}} \right)\overset{\triangle}{ =} {\psi^{lb}}\left( {R_j^d\left[ n \right]} \right),\label{P3_1}
\end{align}
\end{small}
where ${\Sigma _j}\left[ n \right] = \frac{{{C_j}\left[ n \right]\log _2^e}}{{\left( {{C_j}\left[ n \right] + {{\left\| {{{\bf{q}}^r}\left[ n \right] - {{\bf{w}}_j}} \right\|}^2} + {H^2}} \right)\left( {{{\left\| {{{\bf{q}}^r}\left[ n \right] - {{\bf{w}}_j}} \right\|}^2} + {H^2}} \right)}}$ and ${C_j}\left[ n \right] = \frac{{{p_b}{\beta _0}}}{{\sum\limits_{i = 1}^{{K_U}} {x_i^u\left[ n \right]{\beta _0}d_{i,j}^{ - \alpha }{p_i}\left[ n \right] + {\sigma ^2}} }}$.
Similarly, to tackle the non-convexity of $R_i^u\left[ n \right]$, for any given local point ${\left\| {{{\bf{q}}^r}\left[ n \right] - {{\bf{w}}_i}} \right\|^2}$, we have
\begin{small}
\begin{align}
&R_i^u\left[ n \right] \ge {\log _2}\left( {1 + \frac{{{E_i}\left[ n \right]}}{{{{\left\| {{{\bf{q}}^r}\left[ n \right] - {{\bf{w}}_i}} \right\|}^2} + {H^2}}}} \right) - {F_i}\left[ n \right]\times\notag\\
&\left( {{{\left\| {{\bf{q}}\left[ n \right] - {{\bf{w}}_i}} \right\|}^2} - {{\left\| {{{\bf{q}}^r}\left[ n \right] - {{\bf{w}}_i}} \right\|}^2}} \right)\overset{\triangle} {=} {\varphi ^{lb}}\left( {R_i^u\left[ n \right]} \right),\label{P3_2}
\end{align}
\end{small}
where ${F_i}\left[ n \right] = \frac{{\log _2^e}{E_i[n]}}{{\left( {{E_i}\left[ n \right] + {{\left\| {{{\bf{q}}^r}\left[ n \right] - {{\bf{w}}_i}} \right\|}^2} + {H^2}} \right)\left( {{{\left\| {{{\bf{q}}^r}\left[ n \right] - {{\bf{w}}_i}} \right\|}^2} + {H^2}} \right)}}$ and ${E_i}\left[ n \right] = \frac{{{p_i}\left[ n \right]{\beta _0}}}{{{f_b}\left[ n \right] + {\sigma ^2}}}$.
As a result, with \eqref{P3_1} and \eqref{P3_2}, problem $(\rm { P}3)$ can be simplified as
\begin{align}
&\left( {{\rm{\bar P3}}} \right)\mathop {\max }\limits_{\bf{Q}} \sum\limits_{n = 1}^N {\sum\limits_{j = 1}^{{K_D}} {x_j^d\left[ n \right]} } {\psi ^{lb}}\left( {R_j^d\left[ n \right]} \right) +\notag\\
&\qquad\qquad\qquad\qquad\qquad\qquad \sum\limits_{n = 1}^N {\sum\limits_{i = 1}^{{K_U}} {x_i^u\left[ n \right]} } {\varphi ^{lb}}\left( {R_i^u\left[ n \right]} \right)\notag\\
&{\rm s.t.}~~\eqref{const6},\eqref{const7}.\notag
\end{align}
Problem $(\rm { \bar P}3)$ is now a convex optimization problem. Then, we can obtain the locally optimal solution of $(\rm {  P}3)$ by iteratively solving problem $(\rm { \bar P}3)$ to update the UAV trajectory.
\subsection{Stage 4: Uplink user transmit power control}
Finally, we study the uplink user transmit power optimization problem  with  the given UAV trajectory $\bf Q$,  downlink user scheduling ${\bf A}_{\rm D}$, and uplink user scheduling ${\bf A}_{\rm U}$, which is given by
\begin{small}
\begin{align}
\left( {{\rm{P4}}} \right)&\mathop {\max }\limits_{\bf{P}} \sum\limits_{n = 1}^N {\sum\limits_{j = 1}^{{K_D}} {x_j^d\left[ n \right]} } R_j^{d}\left[ n \right] + \sum\limits_{n = 1}^N {\sum\limits_{i = 1}^{{K_U}} {x_i^u\left[ n \right]} } R_i^u\left[ n \right]\notag\\
&{\rm s.t.}~~\eqref{const5}. \notag
\end{align}
\end{small}
The term $R_j^d\left[ n \right]$ is convex w.r.t. $p_i[n]$ which makes problem ${(\rm P}4)$ be a non-convex optimization problem. Similar to that of $(\rm { \bar P}2)$, by taking the first-order Taylor expansion at local point $\{p_i^r[n]\}$, we can approximate it as its lower bound as follows
\begin{small}
\begin{align}
&R_j^d\left[ n \right] \ge {\phi ^{lb}}\left( {R_j^d\left[ n \right]} \right)\overset{\triangle}{ =} {\log _2}\left( {1 + \frac{{{p_b}{h_{b,j}}\left[ n \right]}}{{G_j^r\left[ n \right]}}} \right) - \notag\\
&\qquad\sum\limits_{i = 1}^{{K_U}} {\frac{{{\beta _0}x_i^u\left[ n \right]d_{i,j}^{ - \alpha }{p_b}{h_{b,j}}\left[ n \right]\log _2^e}}{{G_j^r\left[ n \right]\left( {G_j^r\left[ n \right] + {p_b}{h_{b,j}}\left[ n \right]} \right)}}\left( {{p_i}\left[ n \right] - p_i^r\left[ n \right]} \right)},\label{P4_1}
\end{align}
\end{small}
where $G_j^r\left[ n \right] = \sum\limits_{i = 1}^{{K_U}} {x_i^u\left[ n \right]{\beta _0}d_{i,j}^{ - \alpha }p_i^r\left[ n \right] + {\sigma ^2}}$. Then, with \eqref{P4_1}, problem $(\rm {  P}4)$ is simplified as
\begin{small}
\begin{align}
\left( {{\rm{\bar P4}}} \right)&\mathop {\max }\limits_{\bf{P}} \sum\limits_{n = 1}^N {\sum\limits_{j = 1}^{{K_D}} {x_j^d\left[ n \right]} } {\phi ^{lb}}\left( {R_j^d\left[ n \right]} \right) + \sum\limits_{n = 1}^N {\sum\limits_{i = 1}^{{K_U}} {x_i^u\left[ n \right]} } R_i^u\left[ n \right]\notag\\
&{\rm s.t.}~~\eqref{const5}. \notag
\end{align}
\end{small}
Problem $(\rm { \bar P}4)$ is a convex optimization problem. Then, $(\rm { P}4)$ can be approximately solved by successively updating the uplink user transmit power obtained from problem $(\rm { \bar P}4)$.
\subsection{Overall algorithm}
Based on the above  four-stage  sub-problems, we  optimize the four-stage sub-problems in an iterative way, which is summarized in Algorithm~\ref{alg1}. Note that Algorithm~\ref{alg1} is guaranteed to converge a local solution, which can be found  in \cite{wu2018Joint},\cite{spectrum2018wang}.  At last, the continuous user scheduling variable is reconstructed into  binary one  by adopting the following simple criteria \cite{power2018meng}: $x = \left\{ \begin{array}{l}
1,\;\;\;{\rm{if}}\;\;x \ge 0.5,\\
0,\;\;\;{\rm{if}}\;\;x < 0.5,
\end{array} \right.$ where $x \in \left\{ {x_i^u\left[ n \right],x_j^d\left[ n \right],\forall i,j,n} \right\}$.
\begin{small}
\begin{algorithm}[H]\label{alg1}
\caption{Alternating optimization  for problem $(\rm P)$}
\label{alg1}
\begin{algorithmic}[1]
\STATE  \textbf{Initialize}   ${{\bf{q}}^r[n]}$,  $x_i^{u,r}[n]$, $p_i^r[n]$, and set $r \leftarrow 0$  as well as tolerance $\epsilon > 0$.
\STATE  \textbf{repeat}.
\STATE  \quad Solve  $(\rm P1)$ for  given  $\{{\bf q}^r[n],x_i^{u,r}[n],p_i^r[n]\}$, and \\
\quad denote the optimal solution as   $\{{x_j^{d,r+1}\left[ n \right]}\}$.
\STATE  \quad Solve  $(\rm P2)$ for  given   $\{{\bf q}^r[n],{x_j^{d,r+1}\left[ n \right]},p_i^r[n]\}$, and \\
\quad  denote the optimal solution as   $\{x_i^{u,r+1}[n]\}$.
\STATE  \quad Solve  $(\rm P3)$ for  given  $\{x_i^{u,r+1}[n],{x_j^{d,r+1}\left[ n \right]},p_i^r[n]\}$, and \\
\quad    denote the optimal solution as  $\{{\bf q}^{r+1}[n]\}$. \\
\STATE  \quad Solve  $(\rm P4)$ for  given  $\{x_i^{u,r+1}[n],{x_j^{d,r+1}\left[ n \right]},{\bf q}^{r+1}[n]\}$,\\
\quad    and  denote the optimal solution as  $\{p_i^{r+1}[n]\}$. \\
\STATE \quad $r \leftarrow r + 1$.
\STATE \textbf{until} the fractional increase of the objective value of $\left( {{\rm{ P}}} \right)$ is less than tolerance  $\epsilon$.
\end{algorithmic}
\end{algorithm}
\end{small}
It is worth pointing out that all the sub-problems $\rm (P1)$, $\rm (\bar P2)$, $\rm (\bar P3)$, and $\rm (\bar P4)$ are convex, thus the computational complexity of Algorithm  is
${\cal O}\Big ( {{L_4}} \left( {{{\left( {{K_D}N} \right)}^{3.5}} + {L_1}{{\left( {{K_U}N} \right)}^{3.5}} + {L_2}{{\left( {2N} \right)}^{3.5}} + } \right.$ $\left. {\left. {{L_{3}}{{\left( {{K_U}N} \right)}^{3.5}}} \right)} \right)$ with $L_1$, $L_2$, $L_3$, and $L_4$  being the iterative numbers.


\section{Simulation Results}
\begin{figure}
\begin{minipage}[t]{0.5\linewidth}
\centering
\includegraphics[width=1.8in]{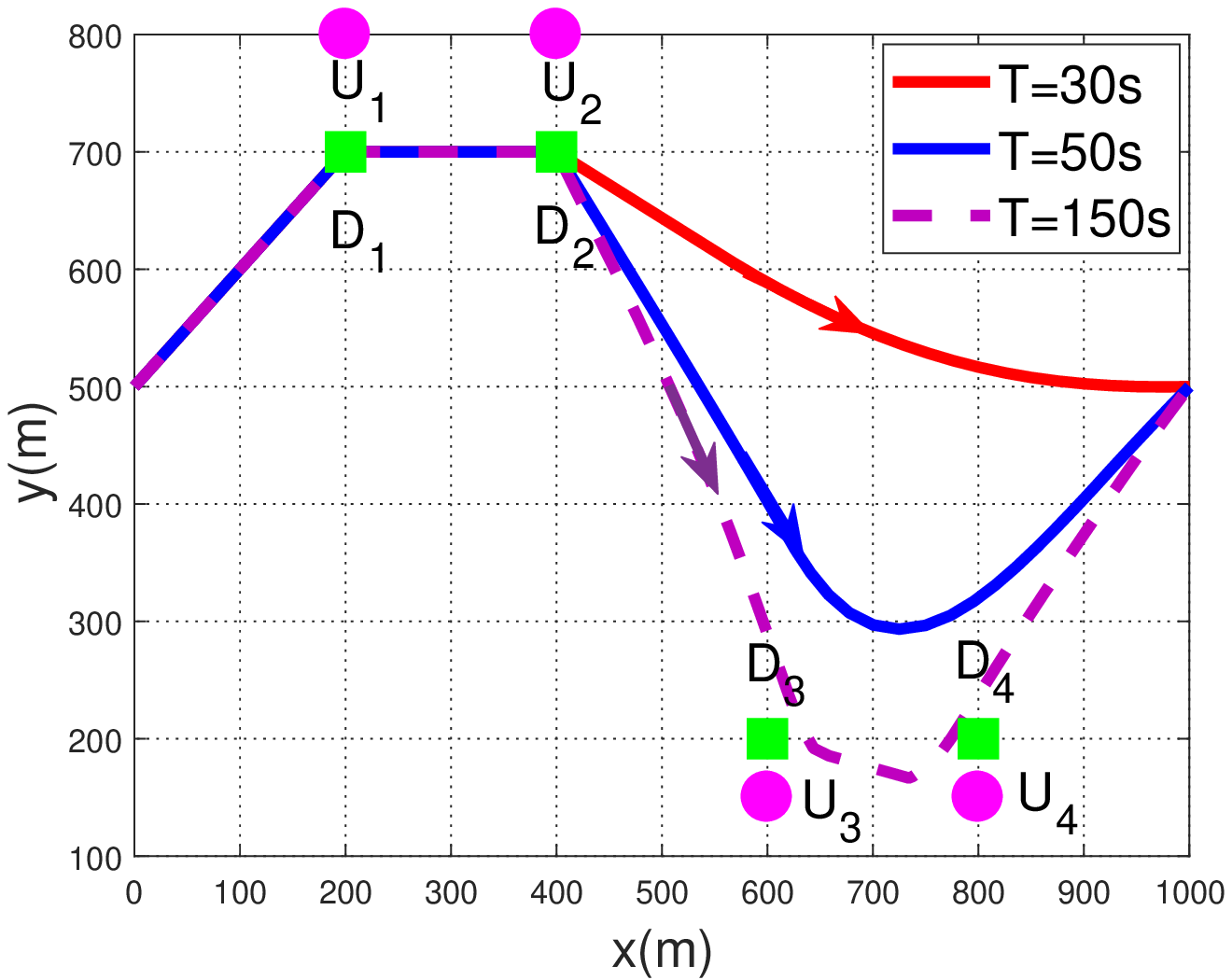}
 \centerline{\scriptsize Fig. 2  UAV trajectory for different period $T$. }
\end{minipage}%
\begin{minipage}[t]{0.5\linewidth}
\centering
\includegraphics[width=1.8in]{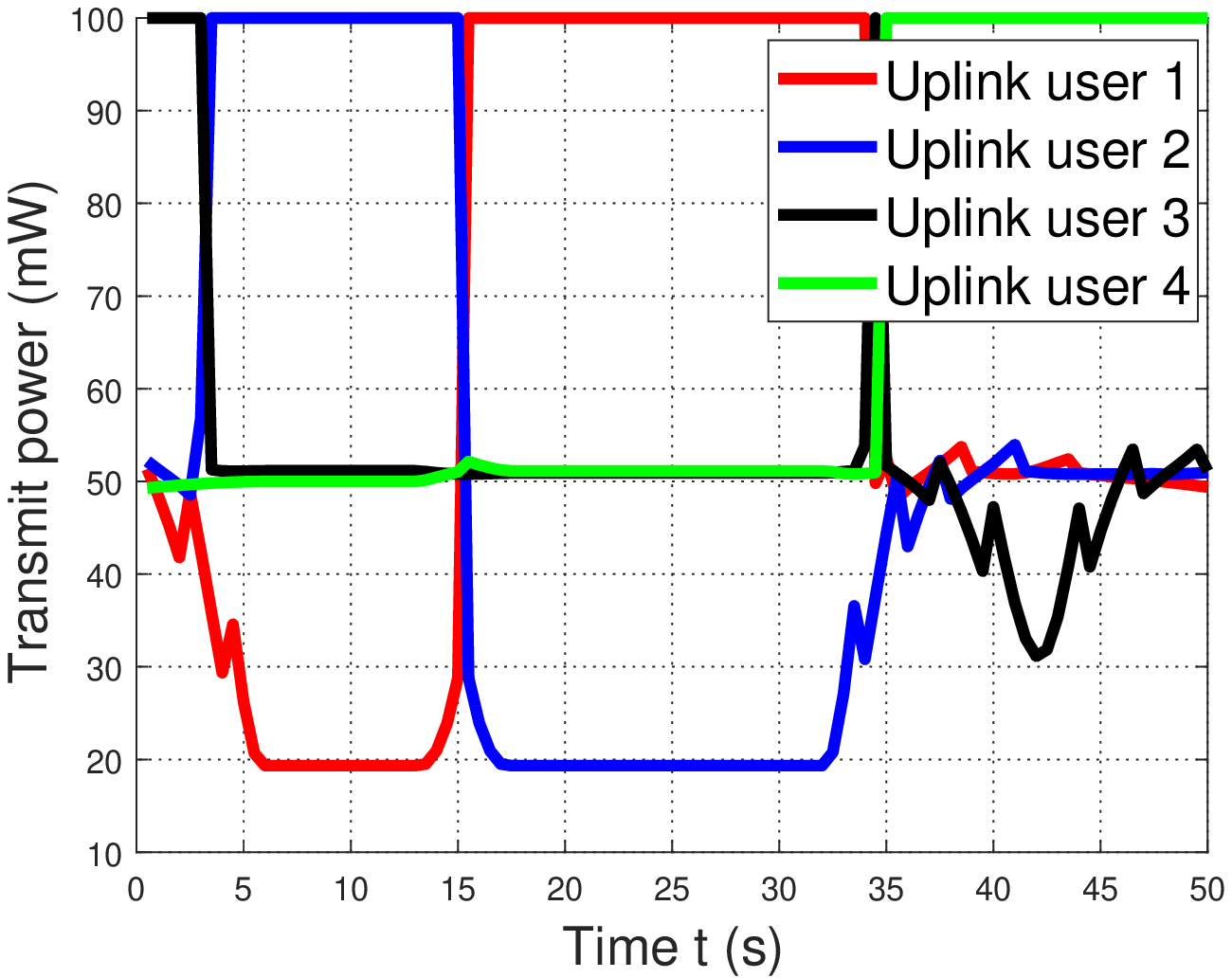}
\centerline{\scriptsize  Fig. 3 Uplink user transmit power.}
\end{minipage}
\end{figure}
In this section, we  evaluate the performance of full-duplex UAV system by our proposed algorithm.  In our example, we consider four downlink users and four uplink users, i.e., $K_D=4$ and  $K_U=4$. The channel gain of the system and noise power are respectively set as ${\beta _0} =  - 60{\rm{dB}}$ and $\sigma^2 =-110{\rm dBm}$ \cite{wu2018Joint}. The system bandwidth is assumed to be $B=1{\rm MHz}$. The UAV  maximum transmit power and speed are respectively  assumed to be  $p_b=0.1{\rm W}$ and $V_{\rm max } = 50{\rm{m/s}}$  \cite{zeng2017energy}. We set  $\alpha=3$,  $\delta=0.5s$,  $P_{\rm max}=0.1{\rm W}$. Unless otherwise specified,  we set  ${H = 100\rm{m}}$ and $f_b[n]=-130{\rm dB}$, $\epsilon=10^{-3}$.

Fig.~2 shows the obtained  UAV trajectories by  our proposed  Algorithm for three different periods, i.e.,  $T=30s$, $T=50s$, and $T=150s$. The UAV's initial and final location are ${\bf q}_{I}=\left( {0,500{\rm m}} \right)^T$ and ${\bf q}_{ F}=\left( {1000{\rm m},500{\rm m}} \right)^T$, respectively. The circle and  square represent locations of uplink  and downlink users, respectively. $D_i$ and $U_i$ denotes the $i$\text{-}th downlink and uplink user, respectively. It is observed that the UAV prefers moving closer to the downlink users rather than the uplink users for all the three different periods.  The reason  is that the downlink users are exposed to the strong interference from the uplink users,  and the UAV moving  closer to the downlink users can  enlarge the downlink capacity.  To illustrate it clearly,  the  uplink user transmit power for $T=50s$ is plotted in Fig.~3. We can observe from this figure that  for the proximity of two uplink users,  one transmits with the maximal power and the other transmits with a lower power  in order to reduce the interference to the nearby downlink users.

\begin{figure}
\begin{minipage}[t]{0.5\linewidth}
\centering
\includegraphics[width=1.8in]{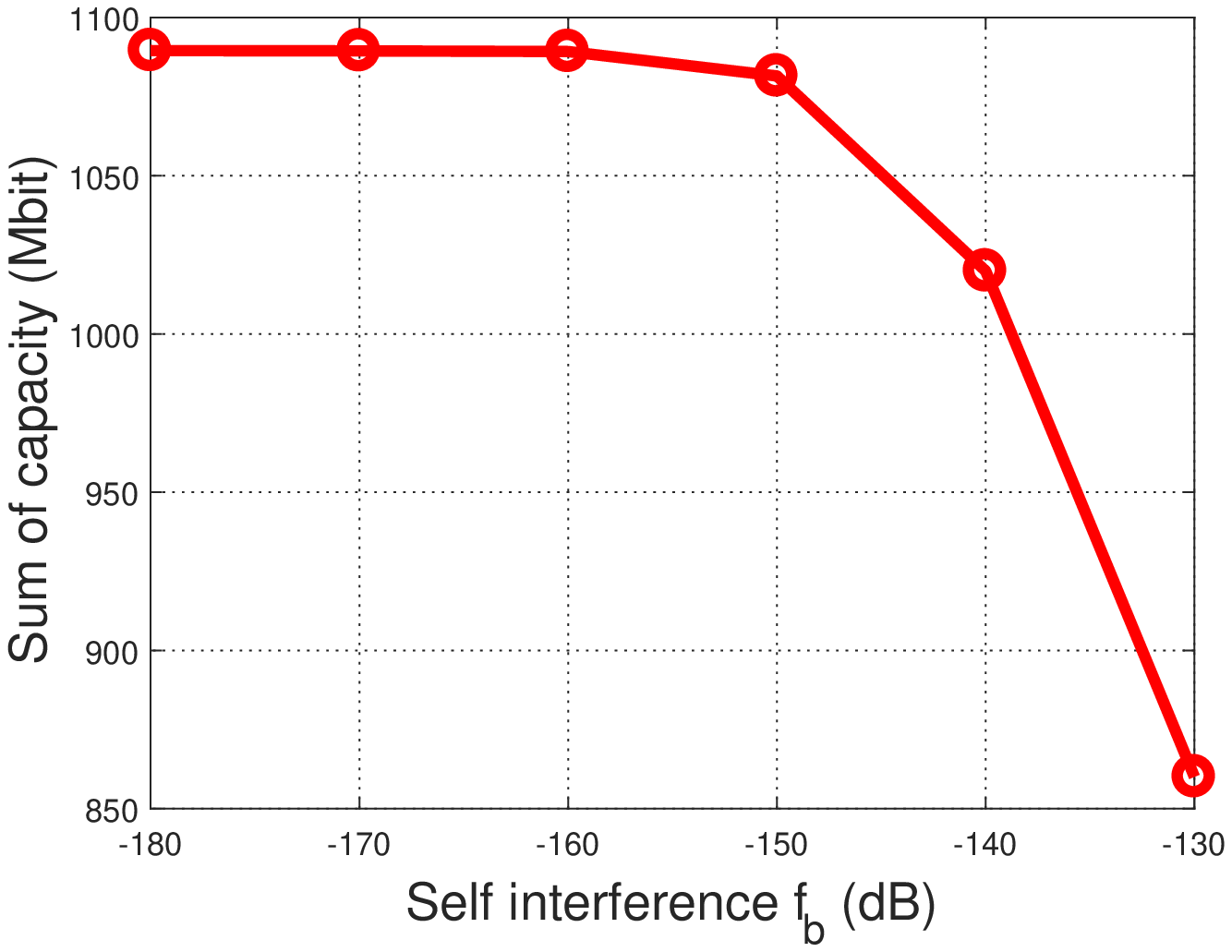}
 \centerline{\scriptsize Fig. 4 Self-interference versus capacity. }
\end{minipage}%
\begin{minipage}[t]{0.5\linewidth}
\centering
\includegraphics[width=1.8in]{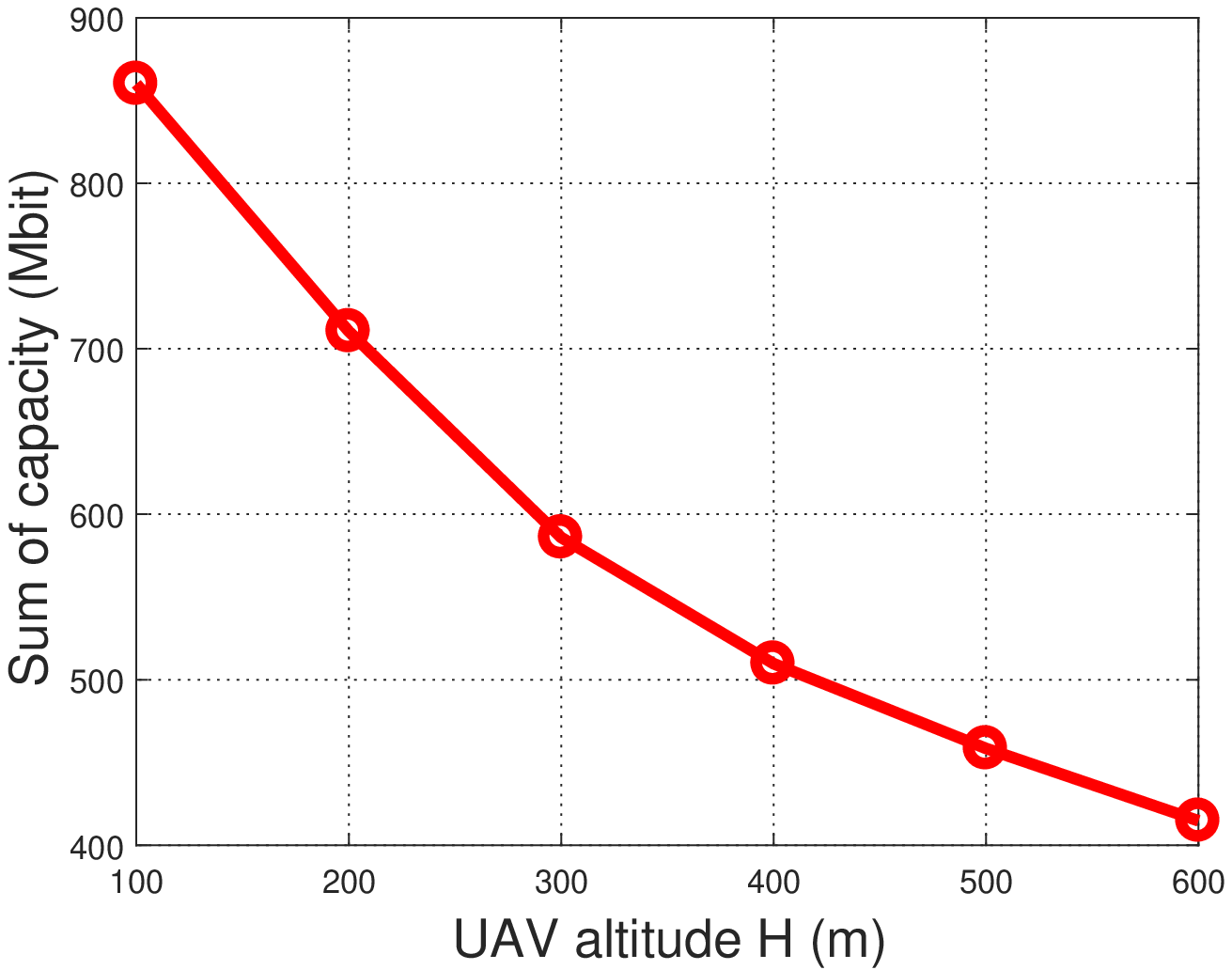}
\centerline{\scriptsize Fig. 5  UAV altitude versus capacity.}
\end{minipage}
\end{figure}

In Fig.~4, we  investigate the impact of self-interference on the sum of system capacity for period $T=70s$. It is observed that the sum of capacity is monotonically non-increasing with the self-interference. Especially, when the power of self-interference $f_b[n]$ below $-150\rm dB$, the sum of capacity is not be changed. This is because the power level of self-interference is much  smaller than the noise power, i.e., $\sigma^2 =-110{\rm dBm}$, thus the impact of self-interference on the system can be neglected.  However, when the power of self-interference $f_b[n]$ exceeds  $-150\rm dB$, the power of interference can not be ignored compared to the noise power, which results in a poor system performance.

In Fig.~5, the effect of UAV altitude on the system performance is studied. It is observed that the system performance is significantly decreasing with the UAV altitude. This is expected since the higher UAV altitude resulting in a lower signal-to-noise (SNR) for uplink and downlink. This result can also be directly  seen in  expressions \eqref{section2downlink} and \eqref{section2uplink}.

In Fig.~6,  we compare our proposed design with the following benchmarks: 1) Ideal no interference: in this scheme when calculating the objective function value, we ignore the interference from the uplink users to the downlink users. Hence, this scheme can serve as the performance upper bound.  2) Proposed design: we jointly optimize the UAV trajectory, downlink/uplink user scheduling, and uplink transmit power.   3) No power control scheme: we assume that  the uplink users transmit with the  maximal  power, and other variables are still optimized. 4) Straight flight scheme: the UAV flies in a straight line from ${\bf q}_{I}$ to ${\bf q}_{F}$ with constant speed $\frac{{\left\| {{{\bf{q}}_F} - {{\bf{q}}_I}} \right\|}}{T}$. 5) Static scheme: UAV stays at a  position with a fixed UAV altitude $H$ that minimizes  the sum of  distance to all the users, i.e., the horizontal UAV location is calculated from ${\bf{q}}_{static}^{opt} = \mathop {\min }\limits_{{{\bf{q}}_{static}}} \sum\limits_{i = 1}^{{{\cal K}_U} \cup {{\cal K}_D}} {{{\left\| {{{\bf{q}}_{static}} - {{\bf{w}}_i}} \right\|}^2}}$. 6) Half-duplex scheme (HD): UAV operates in a half-duplex mode, the interference imposed on the downlink users is disappeared. Note that each time slot is further divided into two time sub-slots with the same duration ${\delta  \mathord{\left/
 {\vphantom {\delta  2}} \right.
 \kern-\nulldelimiterspace} 2}$, and at each time slot, one time sub-slot is assigned to uplink users and the other is assigned to uplink users.   In addition,  for straight flight and static benchmarks, the downlink/uplink user scheduling and uplink user transmit power are still optimized. Several   insights can be made from  Fig.~6. First, we can observe that the interference from the uplink users to the downlink users severely deteriorates  the system capacity. Second, the system capacity can be prominently improved by designing the UAV trajectory. Third, the half-duplex UAV performs worst over other benchmarks. This result shows that with the help of full-duplex technique, it  provides  much system performance gain compared to  the  half-duplex technique. At last,  we  find that the system capacity gain of our proposed algorithm over the power control scheme is marginal. This is because that the downlink capacity can be improved  by reducing the uplink user transmit power  while the uplink capacity will be  decreased. Therefore,  the sum  of downlink and uplink capacity will not improve too much.  However, if we consider a fairness metric  over the  users, we will show that the uplink user transmit power will significantly impact on the system performance  in the  next figure .
\begin{figure}
\begin{minipage}[t]{0.5\linewidth}
\centering
\includegraphics[width=1.8in]{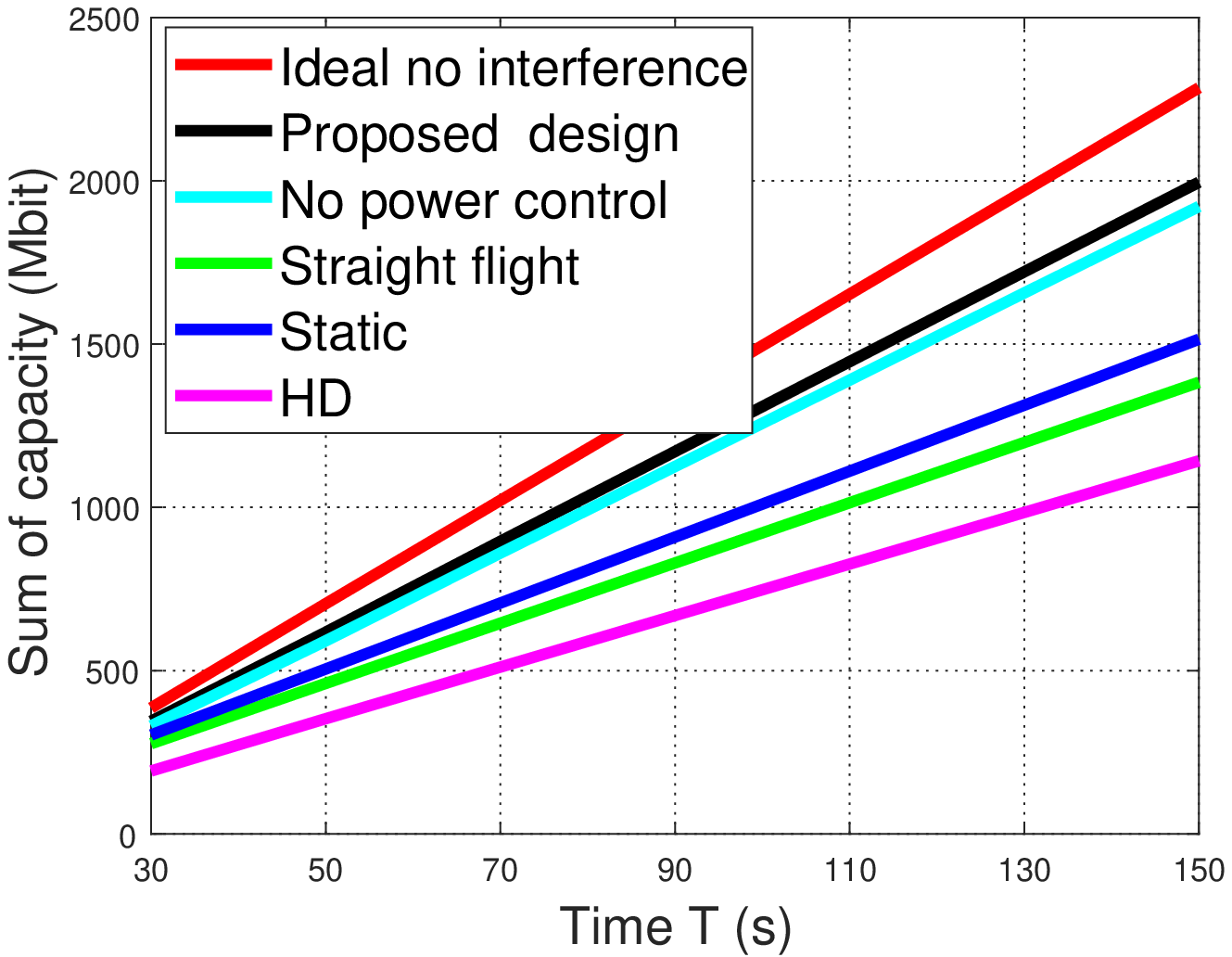}
 \centerline{\scriptsize Fig. 6 Sum of capacity versus $T$. }
\end{minipage}%
\begin{minipage}[t]{0.5\linewidth}
\centering
\includegraphics[width=1.8in]{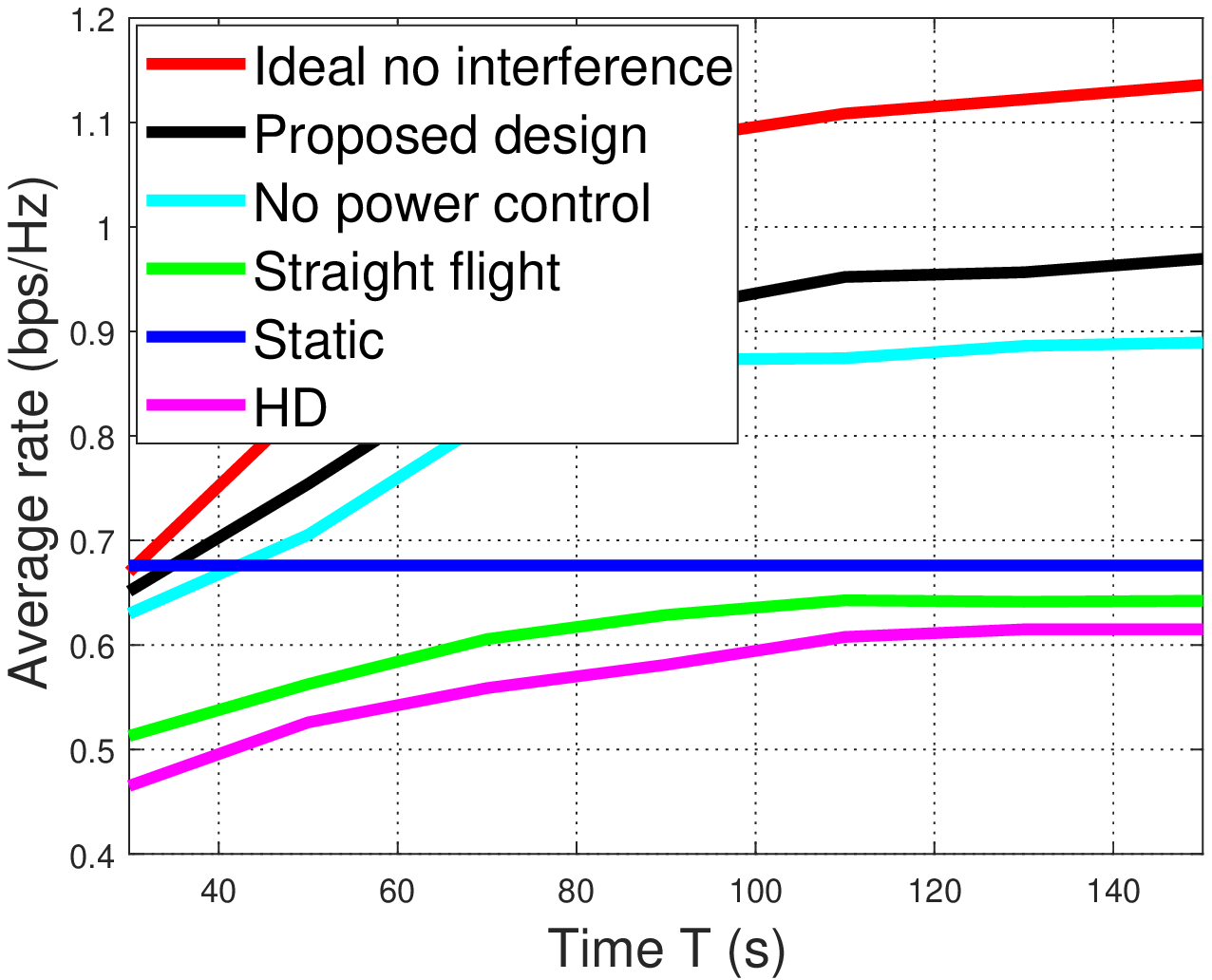}
\centerline{\scriptsize Fig. 7  Max-min rate versus period $T$.}
\end{minipage}
\end{figure}
In Fig.~7, the benchmarks  are the same for that of Fig.~6,  but the goal of this design is to maximize the minimum average capacity  over both uplink users and downlink users for a fair consideration. Similar results can be obtained from that of Fig.~6 except for the last insight. It is observed from Fig.~7 that  the proposed scheme with uplink user power control significantly outperforms no power control scheme in terms of achievable rate. This is because  for achieving a fairness of system performance over the uplink and downlink users, the uplink user transmit should be carefully optimized.


\section{Conclusion}
This paper studied the UAV acted  as a full-duplex base station to serve the ground users. We formulated a sum of uplink and downlink throughput maximization problem  by  alternately optimizing the  UAV trajectory, downlink/uplink user scheduling, and uplink user transmit power. To address the resulting optimization problem, we developed an efficient iterative algorithm to solve it. The simulation results showed that a significant performance gain is improved compared to that of half-duplex UAV.  In addition, the results also showed that the system performance was  prominently improved by optimizing the uplink user transmit power as well as UAV trajectory in terms of average capacity.

%
%

\bibliographystyle{IEEEtran}
\bibliography{TVT20195}
\end{document}